\documentclass{svproc}
\usepackage{url}

\begin{document}
\mainmatter              
\title{Strong Field QED, Astrophysics, and Laboratory Astrophysics}
\titlerunning{Strong Field QED}  
%
\author{Sang Pyo Kim\inst{1},\inst{2}}
\authorrunning{Sang Pyo Kim} 
\tocauthor{Sang Pyo Kim}
\institute{Department of Physics, Kunsan National University, Kunsan 54150, Korea\\
\email{sangkim@kunsan.ac.kr}\\
\and  Asia Pacific Center for Theoretical Physics, Pohang 37673, Korea}

\maketitle              

\begin{abstract}
Astrophysical compact objects, such as magnetars, neutron star mergers, etc, have strong electromagnetic fields beyond the Schwinger field ($B_c = 4.4 \times 10^{13}\, {\rm G}$). In strong electric fields, electron-positron pairs are produced from the vacuum, gamma rays create electron-positron pairs in strong magnetic fields, and propagating photons experience vacuum refringence, etc. Astrophysical compact objects with strong electromagnetic fields open a window for probing fundamental physics beyond weak field QED. Ultra-intense lasers and high-energy charged particles may simulate extreme astrophysical phenomena.
\keywords{magnetar, neutron star merger, QED action, Schwinger pair production, laboratory astrophysics}
\end{abstract}
\section{Introduction}
Two pillars of theoretical physics in the twentieth century are Einstein's general relativity and quantum electrodynamics or quantum field theory. Strong field physics championed by strong gravity and strong QED has emerged as a new field in the twenty-first century. Gravitational waves from mergers of black holes or black holes and neutron stars~\cite{LIGOScientific:2016aoc} and shadows of supermassive black holes have remarkably proved Einstein gravity~\cite{EventHorizonTelescope:2019dse}. Highly magnetized neutron stars, especially magnetars, have been observed to have strong magnetic fields beyond the Schwinger field, the lowest Landau level equaling the rest mass of electrons (MacGill Magnetar Catalog~\cite{Olausen:2013bpa}). The Chandrasekhar and Fermi theoretical bound on the magnetic fields of highly magnetized neutron stars
is $B \leq 10^{18} (M/1.4M_{\odot}) (10 {\rm km}/R)^2\, {\rm G}$~\cite{chandrasekhar1953problems}. The merger of neutron stars has been observed in gravitational waves and electromagnetic waves~\cite{LIGOScientific:2017zic}.

Strong electromagnetic fields make the vacuum polarized, which modifies the Maxwell theory to the nonlinear electrodynamics with Heisenberg-Euler-Schwinger loop correction~\cite{Schwinger1951PhRv...82..664S}. Thus, quantum
processes in supercritical electromagnetic fields differ from those in weak fields, and nonperturbative effects, such as Schwinger pair production, vacuum birefringence, photon-photon scattering, and nonlinear Compton scattering, etc, will be signatures of strong field QED~\cite{Harding:2006qn}. In the laboratory, ultra-intense lasers have been developed with the chirped pulse amplification (CPA) technique, and the current highest intensity of optical lasers is $ I=1.1 \times 10^{23} \, {\rm W/cm^2}$~\cite{Yoon et al.2021Optic...8..630Y}. Using plasma mirrors, the laser intensity may go beyond the critical intensity ($I_{\rm c} = 4.6 \times 10^{29} {\rm W/cm^2} $)~\cite{Thaury et al.2007NatPh...3..424T}, which will open a new window for laboratory astrophysics that tests fundamental physics for astrophysical phenomena in extreme electromagnetic fields~\cite{Kim:2021kif}.

\section{QED One-Loop Effective Action}
Two often used methods for strong field QED action are the worldline formalism and in-out formalism. The one-loop effective action in a background electromagnetic field with gauge potential ${\bf A}$ in the worldline formalism is given by
\begin{eqnarray}
\Gamma [{\bf A}] =  \int_{0}^{\infty} \frac{dT}{T} e^{- m^2 T} \int_{x(0) = x(T)} {\cal D} x(\tau) e^{- \int_0^T d \tau (\dot{\bf x}^2/4 - i q \dot{\bf x} \cdot {\bf A}(x(\tau)))}.
\end{eqnarray}
It is equivalent to all one-loop $n$-photon amplitudes. In fact, Heisenberg and Euler, and Schwinger obtained the one-loop QED action on a constant electromagnetic field in the proper-time integral~\cite{Schwinger1951PhRv...82..664S}:
\begin{eqnarray}
{\cal L}^{(1)} = - \frac{1}{8 \pi^2} \int_{0}^{\infty} ds \frac{e^{-m^2 s}}{s^3} \Bigl[(es)^2 ab \coth(eas) \cot(ebs) - 1 - \frac{(es)^2}{3} (a^2 - b^2) \Bigr],
\end{eqnarray}
where $a$ and $b$ are Lorentz- and gauge-invariants
\begin{eqnarray}
a = \sqrt{\sqrt{{\cal F}^2 + {\cal G}^2} + {\cal F}}, \quad b = \sqrt{\sqrt{{\cal F}^2 + {\cal G}^2} - {\cal F}},
\end{eqnarray}
in terms of the Maxwell scalar and pseudo-scalar of Maxwell tensor $F_{\mu \nu}$:
\begin{eqnarray}
{\cal F} = F^{\mu \nu} F_{\mu \nu}/4, \quad {\cal G} = \epsilon_{\mu \nu \alpha \beta}  F^{\mu \nu} F^{\alpha \beta}/4.
\end{eqnarray}
QED action is invariant under ${\cal L}^{(1)} (\pm a, \pm b) = {\cal L}^{(1)} (a,b)$ due to the CP invariance, and ${\cal L}^{(1)} (a,b) = {\cal L}^{(1)} (i b,-i a)$ due to the E-B duality.
When ${\cal G} = - {\bf E} \cdot {\bf B} = 0\, (b = 0, a = \sqrt{{\bf B}^2 - {\bf E}^2})$, the one-loop action can be written as $(\bar{a} = m^2/2ea )$~\cite{Kim:2021kif}:
\begin{eqnarray}
{\cal L}^{(1)} (a,0) = \frac{m^4}{8 \pi^2 \bar{a}^2} \Bigl[\zeta' (-1, \bar{a})- \frac{1}{12} + \frac{\bar{a}^2}{4} - \Bigl(
\frac{1}{12} - \frac{\bar{a}}{2} + \frac{\bar{a}^2}{2} \Bigr) \ln (\bar{a}) \Bigr],
\end{eqnarray}
and ${\cal L}^{(1)} (a,b)$ can be expanded as a power series of $b$.

In the in-out formalism via the Schwinger variational principle, the scattering matrix gives the one-loop effective action~\cite{Dewitt2003global}
\begin{eqnarray}
e^{i W^{(1)}} = e^{i \int \sqrt{-g} d^4x {\cal L}^{(1)}} = \langle 0, {\rm out} \vert 0, {\rm in} \rangle,
\end{eqnarray}
and the vacuum persistence amplitude
\begin{eqnarray}
\vert \langle 0, {\rm out} \vert 0, {\rm in} \rangle \vert^2 = e^{- 2 {\rm Im} W^{(1)}},\quad 2 {\rm Im} W^{(1)}= \pm {\cal D} \sum_{\kappa} (1 \pm N_{\kappa}).
\end{eqnarray}
Here, ${\cal D}$ denotes the density of states, and ${\cal N}_{\kappa}$ is the mean number of pairs with quantum number $\kappa$.
The particle and antiparticle operators in the in-state and the out-state are related through Bogoliubov transformation
\begin{eqnarray}
a_{\kappa, {\rm out}} = \alpha_{\kappa} a_{\kappa, {\rm in}} + \beta_{\kappa}^* b_{\kappa, {\rm in}}^{\dagger}, \quad b_{\kappa, {\rm out}} = \alpha_{\kappa} b_{\kappa, {\rm in}} + \beta_{\kappa}^* a_{\kappa, {\rm in}}^{\dagger}.
\end{eqnarray}
The mean number of produced pairs and the Bogoliubov relation (upper sign for bosons/lower sign for fermions here and hereafter) are
\begin{eqnarray}
{\cal N}_{\kappa} = \vert \beta_{\kappa} \vert^2, \quad \vert \alpha_{\kappa} \vert^2 \mp \vert \beta_{\kappa} \vert^2 = 1.
\end{eqnarray}
In fact, a boson has the out-vacuum that consists of the multi-particle and antiparticle entangled states of in-vacuum particles
\begin{eqnarray}
\vert 0, {\rm out} \rangle = \prod_{\kappa} \frac{1}{\alpha_{\kappa}} \sum_{n_{\kappa}} \Bigl(- \frac{ \beta_{\kappa}^*}{\alpha_{\kappa}} \Bigr)^{n_{\kappa}} \vert n_{\kappa}, \bar{n}_{\kappa}, {\rm in} \rangle,
\end{eqnarray}
and so does a fermion due to the Pauli blocking
\begin{eqnarray}
\vert 0, {\rm out} \rangle = \prod_{\kappa} \Bigl( - \beta_{\kappa}^* \vert 1_{\kappa}, \bar{1}_{\kappa}, {\rm in} \rangle + \alpha_{\kappa} \vert 0_{\kappa}, \bar{0}_{\kappa}, {\rm in} \rangle \Bigr).
\end{eqnarray}

In the in-out formalism the one-loop effective action with the upper (lower) sign for bosons (fermions) was introduced for at the zero-temperature~\cite{Kim:2008yt,Kim:2011cx}
\begin{eqnarray}
W^{(1)} = - i \ln \langle 0, {\rm out} \vert 0, {\rm in} \rangle = \pm i \sum_{\kappa} \ln (\alpha_{\kappa}^*),
\end{eqnarray}
and at a finite temperature~\cite{Kim:2010qq}
\begin{eqnarray}
e^{i W^{(1)} (T) } = e^{i \int \sqrt{-g} d^4 x {\cal L}^{(1)} (T)} = \langle 0, \beta, {\rm in} \vert U^{\dagger} \vert 0, \beta, {\rm in} \rangle = \frac{{\rm Tr} (U^{\dagger} \rho_{\rm in})}{\rho_{\rm in}},
\end{eqnarray}
where $U$ is the evolution operator, and $\beta$ denotes the inverse temperature. The computation can be performed using the {\it thermal vacuum} as if the zero-temperature.
Setting $\alpha_{\kappa} := e^{-\beta z_{\kappa}}$, the effective action at $T$ per unit four volume is
\begin{eqnarray}
{\cal L}^{(1)} (T) = \pm i {\cal D} \sum_{\kappa, \sigma} \Bigl[ \ln (1 \mp e^{- \beta (\omega_{\kappa} - z_{\kappa})}) - \beta z_{\kappa} - \ln (1 \mp e^{-\beta \omega_{\kappa}}) \Bigr],
\end{eqnarray}
where the first and the second subtractions are the vacuum effective action and the zero field. The purely thermal part of the effective action in a constant electric field
\begin{eqnarray}
\Delta {\cal L}^{(1)} (T, E) = \pm i {\cal D} \sum_{\kappa, \sigma} \Bigl[ \ln (1 \mp e^{- \beta (\omega_{\kappa} - z_{\kappa})}) - \ln (1 \mp e^{-\beta \omega_{\kappa}}) \Bigr],
\end{eqnarray}
gives the imaginary part responsible for pair production
\begin{eqnarray}
2 {\rm Im} \Delta {\cal L}^{(1)}  = \mp {\cal D} \sum_{\kappa, \sigma} \sum_{j =1} \frac{(\pm n_{\rm BE/FD} (\kappa))^{j}}{j} \bigl[(e^{\beta z_{\kappa} } - 1)^j + (e^{\beta z^*_{\kappa} } - 1)^j \bigr],
\end{eqnarray}
as well as the real part (vacuum polarization)
\begin{eqnarray}
{\rm Re} \Delta {\cal L}^{(1)} = \pm {\cal D} \sum_{\kappa, \sigma} \tan^{-1} \Bigl[ \frac{ \sin({\rm Re} {\cal L}^{(1)} (0, \kappa) )}{e^{\beta \omega_{\kappa}} (1 \pm |\beta_{\kappa}|^2)^{(1+ 2 |\sigma|)/2} \mp \cos ({\rm Re} {\cal L}^{(1)} (0, \kappa)) } \Bigr].
\end{eqnarray}
The imaginary part of the effective action in the limit of small mean number of pairs is approximately given by
\begin{eqnarray}
2 {\rm Im} \Delta {\cal L}^{(1)} (T,E) \approx \mp \sum_{\kappa, \sigma} |\beta_{\kappa}|^2 n_{\rm BE/FD} (\kappa)
\end{eqnarray}
and consistent with the pair-production rate~\cite{Kim:2008em}
\begin{eqnarray}
{\cal N}^{\rm spn/sc} (T) = \sum_{\kappa} |\beta_{\kappa}|^2 \coth (\beta \omega_{\kappa}/2) \, \, {\rm or} \, \, \tanh (\beta \omega_{\kappa}/2).
\end{eqnarray}

\section{Vacuum Polarization and Linear Response}

The Heisenberg-Euler-Schwinger QED action and most top-down nonlinear actions belong to the Plebanski class action~\cite{Sorokin:2021tge}
\begin{eqnarray}
{\cal L}_{\rm P} = {\cal L} ({\cal F}, {\cal G}),
\end{eqnarray}
for any analytical function ${\cal L}$ of ${\cal F}$ and $ {\cal G}$.
The polarization and magnetization due to the vacuum polarization
\begin{eqnarray}
{\bf D} = {\bf E} + {\cal P} = \frac{\delta {\cal L}}{\delta {\bf E}}, \quad {\bf H} = {\bf B} - {\cal M} = - \frac{\delta {\cal L}}{\delta {\bf B}},
\end{eqnarray}
give the permittivity, permeability tensors and remarkably the magneto-electric response
\begin{eqnarray}
\delta {\bf D} = \epsilon_{\bf E} \delta {\bf E} + \epsilon_{\bf B} \delta {\bf B}, \quad \delta {\bf H} = \bar{\mu}_{\bf B} \delta {\bf B} + \bar{\mu}_{\bf E} \delta {\bf E}.
\end{eqnarray}
Some magnetoelectric material or multiferroic exhibits the magneto-electric response~\cite{Eerenstein et al.2006Natur.442..759E}. The linear response explains the vacuum birefringence and polarization vectors of a low-energy probe photon~\cite{Kim:2021kif}.

The light modes, refractive indices, and polarization vectors are determined by the matrix (tensor) in dyadics~\cite{Kim:2022lvn}
\begin{eqnarray}
{\bf \Lambda} = - {\cal L}_{\cal F} ((1- n^2) {\bf I} + {\bf n} {\bf n}) + {\cal L}_{\cal FF} {\bf P} {\bf P} +{\cal L}_{\cal FG} ({\bf P} {\bf Q}+ {\bf Q} {\bf P})  + {\cal L}_{\cal GG} {\bf Q} {\bf Q},
\end{eqnarray}
where ${\cal L}_{\cal F}= \partial_{\cal F} {\cal L}$ etc, the wave vector ${\bf k} = \omega {\bf n}$, and
\begin{eqnarray}
{\bf P} = {\bf E} + {\bf n} \times {\bf B}, \quad {\bf Q} = {\bf B} - {\bf n} \times {\bf E}.
\end{eqnarray}
The refractive index ${\rm n} = |{\bf n}|$ is determined by ${\rm det} ({\bf \Lambda}) = 0$. The polarization vectors are the zero-eigenvalue eigenvectors of ${\bf \Lambda} \delta {\bf E} = 0$, which are simplified by writing ${\bf \Lambda}$ in self-conjugate dyadics
\begin{eqnarray}
{\bf \Lambda} = {\bf X} {\bf X} + {\bf Y} {\bf Y} + {\bf Z} {\bf Z} + {\bf U} {\bf U},
\end{eqnarray}
where
\begin{eqnarray}
{\bf X} = a {\bf P} + b {\bf Q}, \quad {\bf Y} = c {\bf Q}, \quad {\bf Z} = d {\bf n}, \quad {\bf U} = e {\bf I}
\end{eqnarray}
with $a^2 = {\cal L}_{\cal FF}$, $b^2 = {\cal L}_{\cal FG}^2/{\cal L}_{\cal FF}$, $c^2 = ({\cal L}_{\cal FF}{\cal L}_{\cal GG} - {\cal L}_{\cal FG}^2)/{\cal L}_{\cal FF}$, $d^2 = -{\cal L}_{\cal F}$, and $e^2 = - {\cal L}_{\cal F} (1-n^2)$.
Then, the polarization vectors can be expressed in terms of the reciprocal vectors as
\begin{eqnarray}
\delta {\bf E} = p {\bf X}' + q {\bf Y}' + r {\bf Z}',
\end{eqnarray}
where, with the volume $V := {\bf X} \times {\bf Y} \cdot {\bf Z}$,
\begin{eqnarray}
{\bf X}' = \frac{{\bf Y} \times {\bf Z}}{V}, \quad {\bf Y}' = \frac{{\bf Z} \times {\bf X}}{V}, \quad {\bf Z}' = \frac{{\bf X} \times {\bf Y}}{V}.
\end{eqnarray}
Constraints follow from ${\rm det} ({\bf \Lambda}) = 0$. The details in both the nondegenerate and degenerate cases may be found in~\cite{Kim:2022lvn}.
The polarization vectors and the Stokes vectors are useful means to probe the electromagnetic structure of highly magnetized neutron stars~\cite{Kim:2024lcn}.

\section{Perspective}

Physics in strong field QED differs from that in weak field QED. Strong field QED phenomena, such as pair production, vacuum birefringence, Breit-Wheeler process, and nonlinear Compton scattering, cannot be properly treated with a few Feynman diagrams. Strong electromagnetic fields contribute quantum corrections, whose one-loop is Heisenberg-Euler-Schwinger QED action. Top-down nonlinear electrodynamics actions, such as Born-Infeld action and ModMax action, have been introduced for theoretical reasons, and most nonlinear electrodynamic actions belong to the Plebanski class that is analytical functions of the Maxwell scalar and pseudo-scalar. A low-energy probe photon experiences vacuum birefringence, which makes the Stokes vectors undergo oscillations in strong electromagnetic regions.

Astrophysical compact objects, such as highly magnetized neutron stars, in particular magnetars, magnetized black holes, and mergers of neutron stars, give rise to significant loop corrections, which can be properly treated in Heisenberg-Euler-Schwinger QED action. Thus, these compact stars provide an arena to test strong field QED as well as strong gravity. Precise measurements of X-ray polarimetry from these compact stars, such as IXPE, eXTP, Compton Telescope etc, will test fundamental physics in strong electromagnetic fields.

Ultra-intense lasers, on the other hand, will provide a tool to test laboratory astrophysics. In the near future, QED effects, such as vacuum birefringence and Schwinger pair production of electrons and positrons, will be measured. Recently, nonlinear Compton scattering at optical wavelength has been observed, which is one of the mechanisms for gamma rays. QED plasma that would occur in jets of highly magnetized neutron stars and mergers of neutron stars can be tested. Hundred PW and EW lasers together with ${\rm GeV}$ electron beams will simulate astrophysical phenomena in the extreme electromagnetic environments.

%
%


\begin{thebibliography}{6}

\bibitem{LIGOScientific:2016aoc}
Abbott, B.P. \textit{et al.} [LIGO Scientific and Virgo]: Observation of Gravitational Waves from a Binary Black Hole Merger.
Phys. Rev. Lett. \textbf{116}, 061102 (2016)
doi:10.1103/PhysRevLett.116.061102

\bibitem{EventHorizonTelescope:2019dse}
Akiyama, K. \textit{et al.} [Event Horizon Telescope]:
First M87 Event Horizon Telescope Results. I. The Shadow of the Supermassive Black Hole.
Astrophys. J. Lett. \textbf{875}, L1  (2019)
doi:10.3847/2041-8213/ab0ec7

\bibitem{Olausen:2013bpa}
Olausen, S.A., Kaspi,  V.M.: The McGill Magnetar Catalog.
Astrophys. J. Suppl. \textbf{212}, 6 (2014)
doi:10.1088/0067-0049/212/1/6

\bibitem{chandrasekhar1953problems} Chandrasekhar, S., Fermi, E.:  Problems of gravitational stability in the presence of a magnetic field. Astrophysical Journal \textbf{118}, 116-141 (1953).

\bibitem{LIGOScientific:2017zic}
Abbott, B.P., \textit{et al.} [LIGO Scientific, Virgo, Fermi-GBM and INTEGRAL]: Gravitational Waves and Gamma-rays from a Binary Neutron Star Merger: GW170817 and GRB 170817A.
Astrophys. J. Lett. \textbf{848}, L13  (2017)
doi:10.3847/2041-8213/aa920c

\bibitem{Schwinger1951PhRv...82..664S} Schwinger, J.: On Gauge Invariance and Vacuum Polarization. Phys. Rev. \textbf{82}, 664–679 (1951) doi:10.1103/PhysRev.82.664

\bibitem{Harding:2006qn}
Harding, A.K., Lai, D.: Physics of Strongly Magnetized Neutron Stars.
Rept. Prog. Phys. \textbf{69}, 2631  (2006)
doi:10.1088/0034-4885/69/9/R03

\bibitem{Yoon et al.2021Optic...8..630Y} Yoon, J.W., \textit{et al.}: Realization of laser intensity over 1023 W/cm2. Optica \textbf{8}, 630 (2021)
    doi:10.1364/OPTICA.420520

\bibitem{Thaury et al.2007NatPh...3..424T} Thaury, C., \textit{et al.}: Plasma mirrors for ultrahigh-intensity optics. Nature Physics \textbf{3}, 424–429 (2007)
    doi:10.1038/nphys595


\bibitem{Kim:2021kif}
Kim, C.M., Kim, S.P.: Magnetars as laboratories for strong field QED.
AIP Conf. Proc. \textbf{2874}, 020013 (2024)
doi:10.1063/5.0215939; Vacuum birefringence at one-loop in a supercritical magnetic field superposed with a weak electric field and application to pulsar magnetosphere.
Eur. Phys. J. C \textbf{83}, 104  (2023)
doi:10.1140/epjc/s10052-023-11243-1

\bibitem{Dewitt2003global} DeWitt, B.S.: The global approach to quantum field theory. Oxford University Press (2003)

\bibitem{Kim:2008yt}
Kim, S.P., Lee, H.K., Yoon, Y.: Effective Action of QED in Electric Field Backgrounds.
Phys. Rev. D \textbf{78}, 105013  (2008)
doi:10.1103/PhysRevD.78.105013; Effective Action of QED in Electric Field Backgrounds II. Spatially Localized Fields.
Phys. Rev. D \textbf{82}, 025015  (2010)
doi:10.1103/PhysRevD.82.025015

\bibitem{Kim:2011cx}
Kim, S.P.: QED Effective Action in Magnetic Field Backgrounds and Electromagnetic Duality.
Phys. Rev. D \textbf{84}, 065004  (2011)
doi:10.1103/PhysRevD.84.065004

\bibitem{Kim:2010qq}
Kim, S.P., Lee, H.K., Yoon, Y.: Nonperturbative QED Effective Action at Finite Temperature.
Phys. Rev. D \textbf{82}, 025016  (2010)
doi:10.1103/PhysRevD.82.025016


\bibitem{Kim:2008em}
Kim, S.P., Lee, H.K., Yoon, Y.: Schwinger Pair Production at Finite Temperature in QED.
Phys. Rev. D \textbf{79}, 045024  (2009)
doi:10.1103/PhysRevD.79.045024

\bibitem{Sorokin:2021tge}
Sorokin, D.P.: Introductory Notes on Non-linear Electrodynamics and its Applications.
Fortsch. Phys. \textbf{70}, 2200092  (2022)
doi:10.1002/prop.202200092

\bibitem{Eerenstein et al.2006Natur.442..759E} Eerenstein, W., Mathur, N.D., Scott, J.F.: Multiferroic and magnetoelectric materials.
Nature \textbf{442}, 759–765 (2006)
doi:10.1038/nature05023

\bibitem{Kim:2022lvn}
Kim, C.M., Kim, S.P.: 3+1 formulation of light modes in nonlinear electrodynamics.
Matter Radiat. Extremes \textbf{10}, in press (2025) [arXiv:2210.12890 [gr-qc]]


\bibitem{Kim:2024lcn}
Kim, D.H, Kim, C.M., Kim, S.P.: Quantum refraction effects in pulsar emission.
Mon. Not. Roy. Astron. Soc. \textbf{531}, 2148-2161  (2024)
doi:10.1093/mnras/stae1304; Strong-field QED effects on polarization states in dipole and quadrudipole pulsar emissions,
Eur. Phys. J. C \textbf{84}, 1322  (2024)
doi:10.1140/epjc/s10052-024-13662-0


\end{thebibliography}
\end{document}